\documentclass
[preprint,preprintnumbers,nobibnotes,showkeys,noshowpacs]{revtex4}%
\usepackage{amssymb}
\usepackage{amsmath}
\usepackage{graphicx}
\usepackage{dcolumn}
\usepackage{bm}
\usepackage[justification=raggedright]{caption}
\usepackage{amsfonts}%
\providecommand{\U}[1]{\protect\rule{.1in}{.1in}}
\begin{document}
\title{Ground-state and thermal entanglements in non-Hermitian XY system with real
and imaginary magnetic fields}
\author{Yue Li$^{1}$, Pan-Pan Zhang$^{1,2}$, Li-Zhen Hu$^{1,2}$, Yu-Liang Xu$^{1}$}
\author{Xiang-Mu Kong$^{1}$}
\altaffiliation{Corresponding author. E-mail address: kongxm668@163.com (X.-M. Kong).}

\affiliation{$^{1}$School of Physics and Optoelectronic Engineering, Institute of
Theoretical Physics, Ludong University, Yantai 264025, China}
\affiliation{$^{2}$Department of physics, Beijing Normal University, Beijing 100875, China}
\date{\today }

\begin{abstract}
In this manuscript, we study the non-Hermitian spin-1/2 XY model in the
presence of the alternating, imaginary and transverse magnetic fields. For the
two-site spin system, we solve exactly the energy spectrum and phase diagram,
also calculate the ground-state and thermal entanglements by using the concept
of\ the concurrence. It is found that the two-site concurrence in the
eigenstate which only depends on the imaginary magnetic field $\eta$ is always
equal to one in the region of $\mathcal{PT}$ symmetry, while it decreases with
$\eta$ in the $\mathcal{PT}$-symmetric broken region. Especially, the
concurrence shows the non-analytic behavior at the exceptional point, and the
same is true in the case of the biorthogonal basis, which indicates that the
concurrence can characterize the phase transition in this non-Hermitian
system. The interesting thing is that $\eta$ weakens the thermal entanglement
when the system is isotropic and enhances the entanglement when the system
becomes the Ising model. For the one-dimensional spin chain, the magnetization
and entanglement are further studied by using the two-spin cluster mean-field
approximation. The results show that their variations have opposite trends
with the magnetic fields. Moreover, the system exists the first-order quantum
phase transitions for some anisotropic parameters in the $\mathcal{PT}%
$-symmetry region, and the entanglement changes suddenly at the quantum phase
transition point.

\end{abstract}
\keywords{Entanglement; Concurrence; Non-Hermitian XY system; $\mathcal{PT}$ symmetry;
Exceptional point; Mean-field theory}\maketitle

\section{Introduction}

As we have learned, Hamiltonians are required to be Hermitian in quantum
mechanics, to ensure that the eigenvalues of Hamiltonians are real and the
unitarity of evolution of the system state over time. Hermitian Hamiltonians
generally describe closed systems, while the quantum systems which interact
with the external environment can be represented by equivalent non-Hermitian
Hamiltonians \cite{1}. To explore these kinds of questions, as early as 1943,
W. Pauli proposed the non-Hermitian operator and its theory of self-consistent
inner product, which created a precedent in the study of non-Hermitian quantum
theories \cite{2,3}, and there are plenty of studies on non-Hermitian systems
subsequently \cite{4,5,6,7,8,9}. In 1998, Bender and Boettcher studied the
systems with non-Hermitian Hamiltonians more deeply and found that
non-Hermitian Hamiltonians with parity and time-reversal $\left(
\mathcal{PT}\right)  $ symmetry can still have full real spectrums \cite{10}.
And it made a tremendous contribution to research and development of
non-Hermitian quantum mechanics.

In recent years, many studies have been done on non-Hermitian systems in
theory and experiment and have found interesting phenomena that do not exist
in Hermitian systems. Theoretically, there have been a large number of
researches on the non-Hermitian in skin effect, generalized topological
phases, the new non-Hermitian universalities, and so on
\cite{11,12,13,14,15,16,17}. Experimentally, a lot of work has been studied in
quantum information, quantum optical systems, photonic crystals, mechanical
systems, biological systems, and other fields \cite{18,19,20,21,22,23,24}.

It is well known that quantum entanglement plays an important role in various
fields such as quantum information, condensed matter physics and statistical
physics \cite{25,26,27} and it is a characteristic of quantum systems, which
provides a unique method for exploring the properties of quantum many-body
systems. In particular, it has become a significant concept in condensed
matter physics for characterizing and exploring the phases of matter
\cite{28,29}. For the past few years, a number of inspiring advances have been
made in the study of quantum entanglement and quantum phase transition
\cite{30,31,32,33}. In the fields of black hole physics, holography and
non-equilibrium quantum dynamics, different entanglement measures have also
attracted widespread attention \cite{34,35,36,37}. There is also much work
that has been done on the entanglement properties of the Hermitian spin
systems \cite{38,39,40,41,42,43}. The measurement of entanglement has also
been achieved in experiments, such as the entanglement entropy \cite{44,45}.

Some meaningful and pioneering work has been done on the entanglement
characteristics of non-Hermitian fermion systems, such as non-Hermitian
topology and quantum quenching in non-Hermitian Hamiltonians
\cite{46,47,48,49}. However, there are few studies on the entanglement
properties in non-Hermitian spin systems \cite{YLee,India}. In particular,
there are very few studies using concurrence to measure it. For the spin
systems, the XY systems are widely investigated, for instance, quantum
entanglement, quantum discord, dynamics of quantum coherence, quantum Fisher
information, and other aspects \cite{50,51,52,53}. The interest about XY
systems is due to experimental work on quasicrystals and quasiperiodic
superlattices \cite{54}. We study the ground-state and thermal entanglements
respectively in the non-Hermitian spin-$1/2$ XY system by using concurrence as
an entanglement measurement method and the mean-field theory in this
manuscript. The purpose is to find the influence of the non-Hermitian term on
entanglement and some peculiar properties of the system.

The organizational structure of this manuscript is as follows: In Sec.
\ref{section2}, we introduce the non-Hermitian XY model and discuss the
$\mathcal{PT}$ symmetry of the system. The ground-state phase diagram is
studied in Sec. \ref{section3}. In Sec. \ref{section4} and \ref{sec50}, the
ground-state entanglement is discussed. Sec. \ref{sec5} studies the thermal
entanglement. The magnetization and thermal entanglement are investigated by
mean-field approximation in Sec. \ref{sec6}. We summarize our results in Sec.
\ref{sec7}.%
\begin{figure}
[ptb]
\begin{center}
\includegraphics[
height=2.4344in,
width=5.5486in
]%
{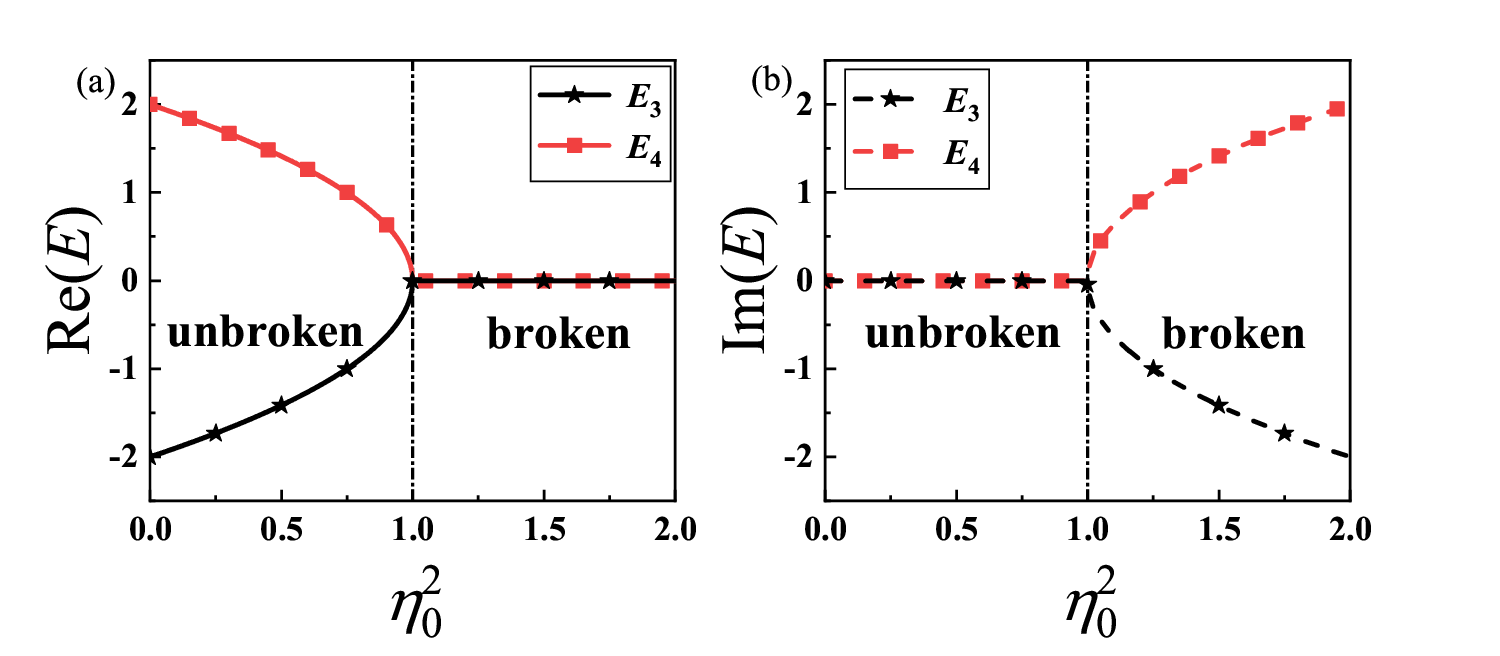}%
\caption{The real and imaginary parts of the eigenvalues $E_{3}$ and $E_{4}$
as the functions of the square of the imaginary magnetic field. (a) the real
parts of $E_{3}$ and $E_{4}$, (b) the imaginary parts of $E_{3}$ and $E_{4}$.
$\eta_{0}^{2}$ $<$ 1 is the $\mathcal{PT}$-symmetric region and $\eta_{0}^{2}$
$>$ 1 is the $\mathcal{PT}$-symmetric broken one. The exceptional point occurs
at $\eta_{0}^{2}$=1.}%
\label{Figure1}%
\end{center}
\end{figure}

\section{model and exceptional point\label{section2}}

The Hamiltonian of the non-Hermitian spin-1/2 XY system in a one-dimensional
lattice is%
\begin{equation}
H=-\frac{J}{2}%
{\displaystyle\sum\limits_{l=1}^{N}}
\left[  \left(  1+\gamma\right)  \sigma_{l}^{x}\sigma_{l+1}^{x}+\left(
1-\gamma\right)  \sigma_{l}^{y}\sigma_{l+1}^{y}\right]  -h%
{\displaystyle\sum\limits_{l=1}^{N}}
\sigma_{l}^{z}+i\eta%
{\displaystyle\sum\limits_{l=1}^{N}}
\left(  -1\right)  ^{l}\sigma_{l}^{z}, \label{equ1}%
\end{equation}
where $\sigma_{l}^{\alpha}(\alpha=x,y,z)$ are the Pauli operators and satisfy
the periodic boundary condition $\sigma_{l}^{\alpha}=\sigma_{l+N}^{\alpha}$,
$N$ is the number of sites (spins) in the system. $J$ is the nearest neighbor
exchange coupling constant ($J>0$ and $J<0$ represent ferromagnetic and
anti-ferromagnetic systems, respectively), $\gamma$ is the anisotropic
parameter, $h$ is the external magnetic field and $i\eta$ an imaginary,
transverse magnetic field. $\eta$ is a real number which measures the
deviation of $H$ from Hermiticity. If $\eta$ is large enough to make complex
some of the eigenvalues of $H$, this symmetry is spontaneously broken
\cite{55}.

In this section, we study the XY model with two sites when $J>0$ whose
Hamiltonian is
\begin{equation}
H=-\frac{J}{2}\left[  \left(  1+\gamma\right)  (\sigma_{1}^{x}\sigma_{2}%
^{x}+\sigma_{2}^{x}\sigma_{1}^{x})+\left(  1-\gamma\right)  (\sigma_{1}%
^{y}\sigma_{2}^{y}+\sigma_{2}^{y}\sigma_{1}^{y})\right]  -h(\sigma_{1}%
^{z}+\sigma_{2}^{z})+i\eta(-\sigma_{1}^{z}+\sigma_{2}^{z}). \label{equ2}%
\end{equation}
The eigenvalues and corresponding eigenstates of the Hamiltonian (\ref{equ2})
are written as
\begin{subequations}
\begin{align}
E_{1}  &  =-2J\sqrt{h_{0}^{2}+\gamma^{2}},\text{ }\left\vert \varphi
_{1}\right\rangle =\frac{1}{\sqrt{\gamma^{2}+d_{1}^{2}}}\left[  d_{1}%
\left\vert \uparrow\uparrow\right\rangle +\gamma\left\vert \downdownarrows
\right\rangle \right]  ,\label{4a}\\
E_{2}  &  =2J\sqrt{h_{0}^{2}+\gamma^{2}},\text{ }\left\vert \varphi
_{2}\right\rangle =\frac{1}{\sqrt{\gamma^{2}+d_{2}^{2}}}\left[  d_{2}%
\left\vert \uparrow\uparrow\right\rangle +\gamma\left\vert \downdownarrows
\right\rangle \right]  ,\label{4b}\\
E_{3}  &  =-2J\sqrt{1-\eta_{0}^{2}},\text{ }\left\vert \varphi_{3}%
\right\rangle =\frac{1}{\sqrt{1+\left\vert d_{3}\right\vert ^{2}}}\left[
d_{3}\left\vert \uparrow\downarrow\right\rangle +\left\vert \downarrow
\uparrow\right\rangle \right]  ,\label{4c}\\
E_{4}  &  =2J\sqrt{1-\eta_{0}^{2}},\text{ }\left\vert \varphi_{4}\right\rangle
=\frac{1}{\sqrt{1+\left\vert d_{4}\right\vert ^{2}}}\left[  d_{4}\left\vert
\uparrow\downarrow\right\rangle +\left\vert \downarrow\uparrow\right\rangle
\right]  , \label{4d}%
\end{align}
where $h_{0}=h/J$ and $\eta_{0}=\eta/J$ are the reduced real magnetic field
and imaginary magnetic field, respectively, as well as
\end{subequations}
\begin{align}
d_{1}  &  =h_{0}+\sqrt{h_{0}^{2}+\gamma^{2}},\text{ }d_{2}=h_{0}-\sqrt
{h_{0}^{2}+\gamma^{2}},\nonumber\\
d_{3}  &  =i\eta_{0}+\sqrt{1-\eta_{0}^{2}},\text{ }d_{4}=i\eta_{0}%
-\sqrt{1-\eta_{0}^{2}}. \label{equ5}%
\end{align}

On the basis of the above statement, it is seen that all the eigenvalues of
$H$ are real if and only if $1-\eta_{0}^{2}\geq0$. Fig. 1 gives the variations
of the real and imaginary parts of $E_{3}$ and $E_{4}$ with $\eta_{0}^{2}$,
respectively. When $\eta_{0}^{2}<1$, the eigenvalues $E_{3}$ and $E_{4}$
remain real and the system is the $\mathcal{PT}$-symmetric. When $\eta_{0}%
^{2}>1$, the both eigenvalues are purely imaginary. At the moment, the system
is in the $\mathcal{PT}$-symmetric broken region. The circle dot $\left(
\text{blue}\right)  $ is the exceptional point $\left(  \text{EP}\right)  $ at
which the real and the imaginary parts of the eigenvalues are both zero and
$E_{3}=E_{4}$ when $\eta_{0}^{2}=1$. Moreover, it also corresponds to the
phase transition point from the $\mathcal{PT}$-symmetric phase to the broken one.

\section{Ground-state Phase Diagram\label{section3}}

On account of the previous discussion, we find that the ground state must be
$\left\vert \varphi_{1}\right\rangle $ or $\left\vert \varphi_{3}\right\rangle
$. In this section, we discuss the conditions in which $\left\vert \varphi
_{1}\right\rangle $ and $\left\vert \varphi_{3}\right\rangle $ are the ground
states respectively or degenerate ground states. Fig. 2 is the ground-state
phase diagram of the system and shows graphically the two possible ground
states when $\gamma=0,0.5$.%
\begin{figure}
[ptb]
\begin{center}
\includegraphics[
height=1.9527in,
width=6.0502in
]%
{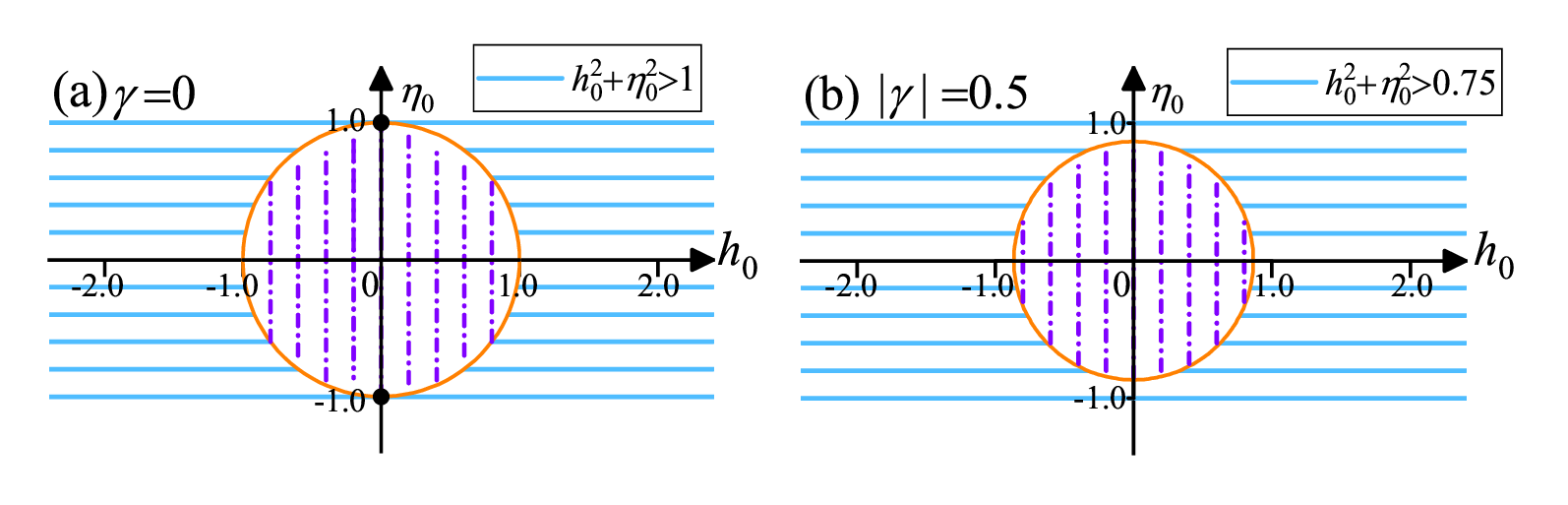}%
\caption{Ground-state phase diagrams of the XY model with two sites. The
shaded areas which are composed of transverse (blue) lines and vertical
(purple) lines correspond to $\left\vert \varphi_{1}\right\rangle $ and
$\left\vert \varphi_{3}\right\rangle $, respectively.}%
\label{Figure2}%
\end{center}
\end{figure}

The ground-state phase diagram in the $h_{0}$-$\eta_{0}$ plane is shown when
$\gamma=0$ in Fig. 2(a) where the shaded area consisting of transverse (blue)
lines (excluding (0,1) and (0,-1)), that is $h_{0}^{2}+\eta_{0}^{2}>1$,
corresponds to the ground state is $\left\vert \varphi_{1}\right\rangle $.
While $h_{0}^{2}+\eta_{0}^{2}<1$, then $\left\vert \varphi_{3}\right\rangle $
is the ground state corresponding to the shaded area composing of vertical
(purple) lines inside the circle. It corresponds to the circle (orange) border
(Not including black circle dots) that $\left\vert \varphi_{1}\right\rangle $
and $\left\vert \varphi_{3}\right\rangle $ are degenerate ground states when
$h_{0}^{2}+\eta_{0}^{2}=1$.

In the case of $0<\left\vert \gamma\right\vert <1$, $\left\vert \varphi
_{3}\right\rangle $ is the ground state, which is still possible. Without loss
of generality, Fig. 2(b) shows the ground-state phase diagram in the $h_{0}%
$-$\eta_{0}$ plane when $\left\vert \gamma\right\vert =0.5$. In the case of
$h_{0}^{2}+\eta_{0}^{2}>0.75$, $\left\vert \varphi_{1}\right\rangle $ is the
ground state which corresponds to the shaded area consisting of blue
(transverse) lines. The shaded area consisting of purple (vertical) lines
inside the circle is the case of $h_{0}^{2}+\eta_{0}^{2}<0.75$, which
corresponds to the ground state is $\left\vert \varphi_{3}\right\rangle $.
When $h_{0}^{2}+\eta_{0}^{2}=0.75$, both $\left\vert \varphi_{1}\right\rangle
$ and $\left\vert \varphi_{3}\right\rangle $ are ground states, which
corresponds to the orange $\left(  \text{circle}\right)  $ border in Fig. 2(b).

When $\left\vert \gamma\right\vert =1$, If real and imaginary magnetic fields
are not zero simultaneously, then $\left\vert \varphi_{1}\right\rangle $ is
the ground state. Otherwise $E_{1}=E_{3}$, the ground states are degenerate.
When $\gamma>1$, $h_{0}^{2}+\eta_{0}^{2}>1-\gamma^{2}$, $\left\vert
\varphi_{1}\right\rangle $ is the ground state certainly.

\section{Ground-state Entanglement\label{section4}}

The previous section has discussed the ground state phase diagram of the
system. The ground-state entanglement of the system is studied by using
concurrence in this section. The concurrence is defined as \cite{56}%
\begin{equation}
C=\max\left[  0,\sqrt{\lambda_{1}}-\sqrt{\lambda_{2}}-\sqrt{\lambda_{3}}%
-\sqrt{\lambda_{4}}\right]  , \label{equ6}%
\end{equation}
where $\lambda_{j}\left(  j=1,2,3,4\right)  $ are the eigenvalues, in
decreasing order, of the non-Hermitian matrix $R=\rho\widetilde{\rho}$. Note
that each $\lambda_{j}$ is a non-negative real number \cite{56}. $\rho$ is the
density matrix, and $\widetilde{\rho}$ is the spin-flipped density matrix
which can be written as $\widetilde{\rho}=\left(  \sigma^{y}\otimes\sigma
^{y}\right)  \rho^{\ast}\left(  \sigma^{y}\otimes\sigma^{y}\right)  $, where
$\rho^{\ast}$ is the complex conjugate of $\rho$.%
\begin{figure}
[ptb]
\begin{center}
\includegraphics[
height=4.0387in,
width=5.047in
]%
{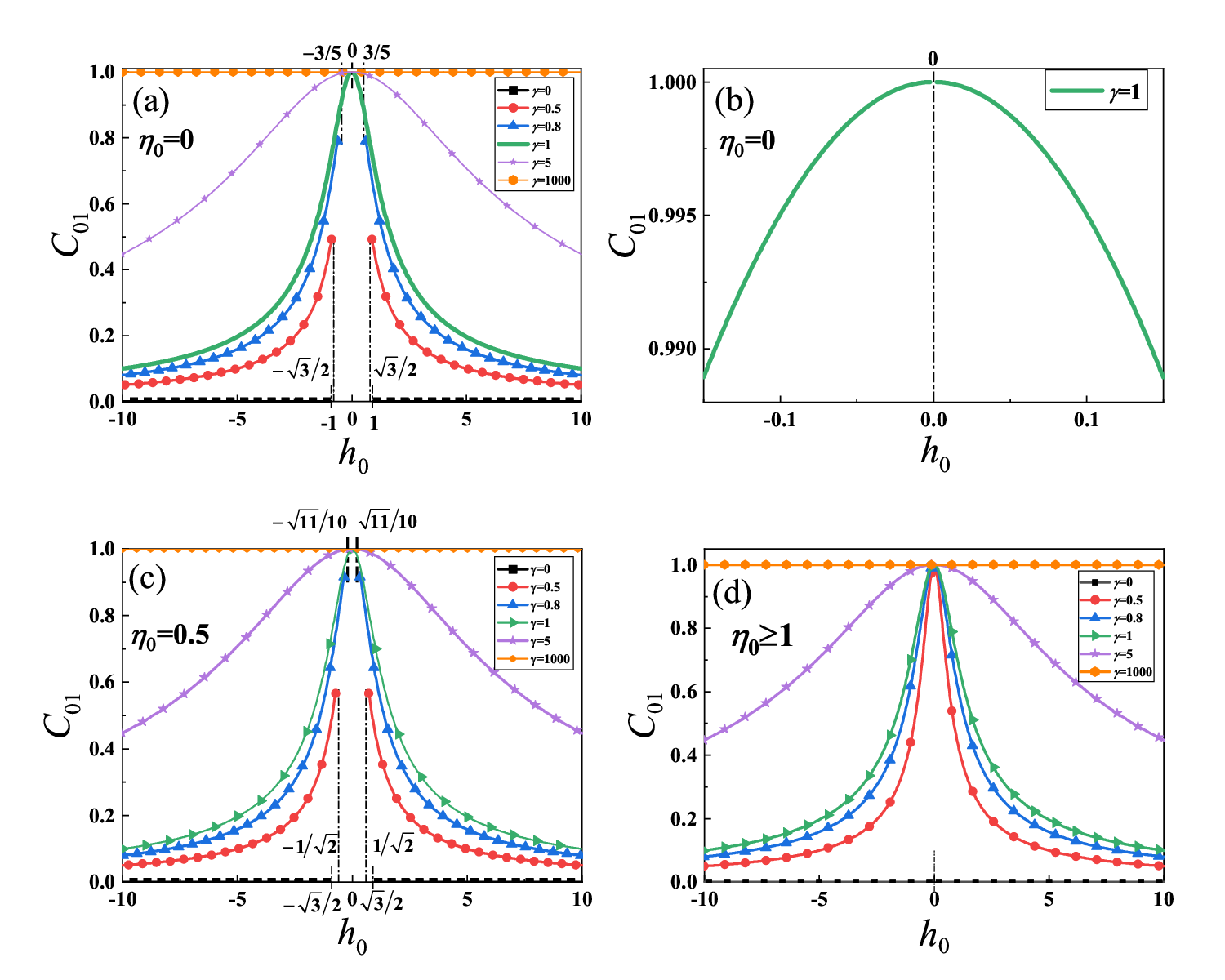}%
\caption{Concurrence $C_{01}$ versus the real magnetic field for different the
anisotropy parameters and imaginary magnetic fields: $($b$)$ the entanglement
behavior in the range of $\left\vert h_{0}\right\vert <0.15$ when $\eta_{0}%
=0$, $\gamma=1$. The range of $h_{0}$ between the marked dashed line and the
dashed line cannot take a value. Concurrence is zero (Square dot line) between
two spins when $\gamma=0$.}%
\label{Figure3}%
\end{center}
\end{figure}

$\left(  \text{\textbf{1}}\right)  $ \textbf{The case of }$h_{0}^{2}+\eta
_{0}^{2}>1-\gamma^{2}$

In this situation, $\left\vert \varphi_{1}\right\rangle $ is the ground state
which only depends on $h_{0}$ and $\gamma$. The ground-state density matrix of
the system is obtained as%
\begin{equation}
\rho_{01}=\left(
\begin{array}
[c]{cccc}%
a_{1} & 0 & 0 & a_{2}\\
0 & 0 & 0 & 0\\
0 & 0 & 0 & 0\\
a_{3} & 0 & 0 & a_{4}%
\end{array}
\right)  , \label{equ7}%
\end{equation}
where%
\begin{equation}
a_{1}=\frac{d_{1}^{2}}{\gamma^{2}+d_{1}^{2}},\text{ }a_{2}=a_{3}=\frac{\gamma
d_{1}}{\gamma^{2}+d_{1}^{2}},\text{ }a_{4}=\frac{\gamma^{2}}{\gamma^{2}%
+d_{1}^{2}}. \label{equ8}%
\end{equation}

Base on the definition, when the ground state is $\left\vert \varphi
_{1}\right\rangle $, the concurrence can be written as%
\begin{equation}
C_{01}=\frac{\left\vert \gamma\right\vert }{\sqrt{h_{0}^{2}+\gamma^{2}}}.
\label{equ9}%
\end{equation}
It can be seen that there is no entanglement between two spins when $\gamma=0$
and $h_{0}\neq0$ within the value ranges of parameters discussed in the
previous section. In this situation, $\left\vert \varphi_{1}\right\rangle
=\left\vert \upuparrows\right\rangle $ is a direct product state, which is
consistent with the above case.

In the case of $0<\left\vert \gamma\right\vert <\infty$, if $h_{0}=0$, then
$C_{01}=1$ and $\left\vert \varphi_{1}\right\rangle =\frac{1}{\sqrt{2}}\left[
\left\vert \upuparrows\right\rangle +\left\vert \downdownarrows\right\rangle
\right]  $, the system is in the Bell state; the entanglement disappears,
which occurs in $\left\vert h_{0}\right\vert \rightarrow\infty$. In addition,
the entanglement between two spins also is the largest, which takes place when
$\left\vert \gamma\right\vert \rightarrow\infty$ and $0\leq\left\vert
h_{0}\right\vert <\infty$.

The above results can be seen more obviously in Fig. 3 which is the variations
of concurrence with $h_{0}$ for different $\gamma$. We also find that the
anisotropic parameter enhances entanglement, while the real magnetic field
weakens entanglement, and $\eta_{0}$ only affects the value range of $h_{0}$.
In the $\mathcal{PT}$-symmetric broken region, the ground state is $\left\vert
\varphi_{1}\right\rangle $ due to the real part of $E_{3}$ is zero and $E_{1}$
is always less than zero when $h_{0}$ and $\gamma$ are not simultaneously zero
\cite{Japan}. Specially, the ground-state concurrence has the maximum when
$h_{0}=0$ at the exceptional point.%
\begin{figure}
[ptb]
\begin{center}
\includegraphics[
height=3.0986in,
width=4.0413in
]%
{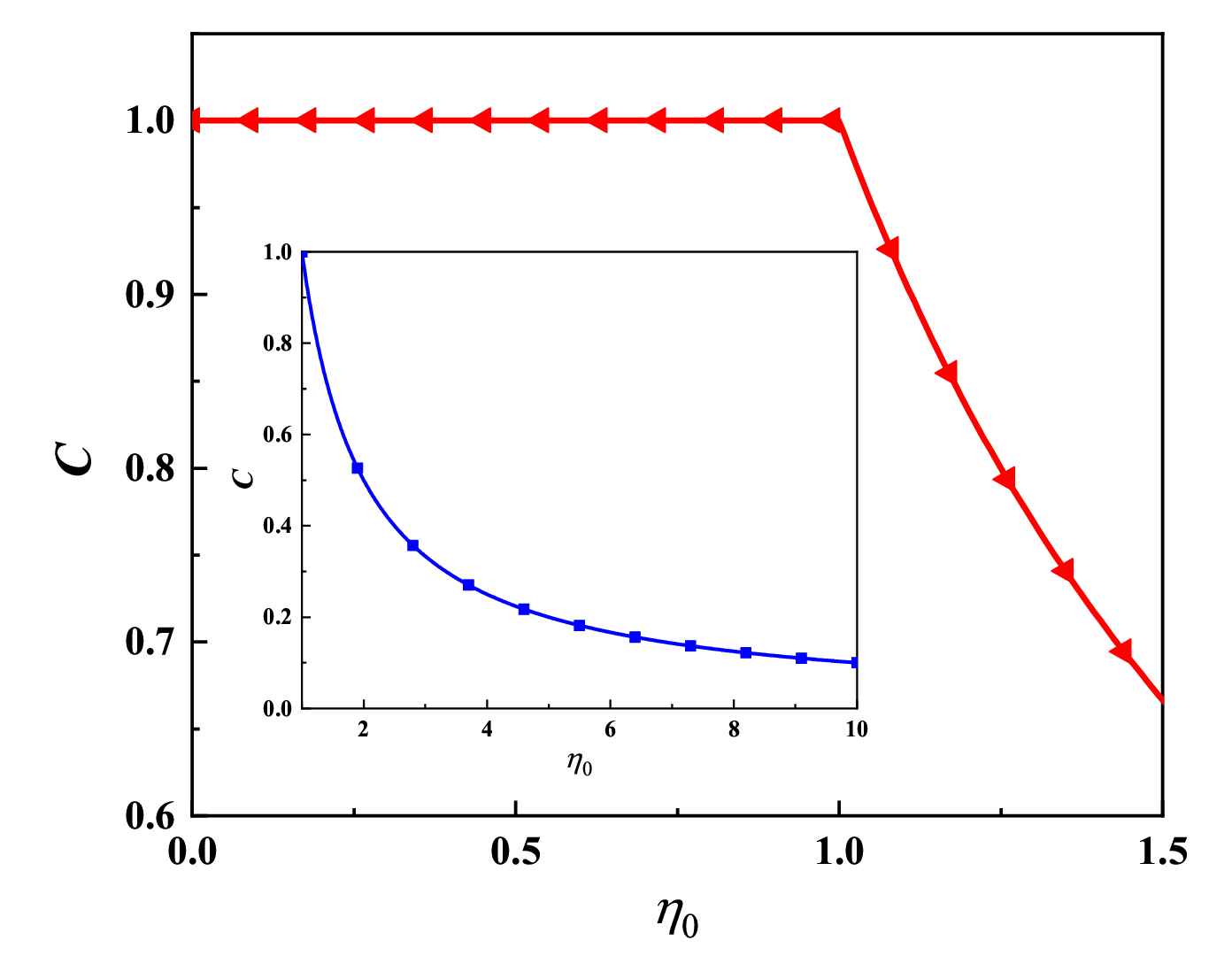}%
\caption{Concurrence versus the imaginary magnetic field. The illustration
shows concurrence when the system is in the $\mathcal{PT}$-symmetry broken
region.}%
\label{Figure4}%
\end{center}
\end{figure}

$\left(  \text{\textbf{2}}\right)  $ \textbf{The case of }$h_{0}^{2}+\eta
_{0}^{2}<1-\gamma^{2}$

In this case, $\left\vert \varphi_{3}\right\rangle $ which only depends on
$\eta_{0}$, is the ground state. The ground-state density matrix of the system
is%
\begin{equation}
\rho_{02}=\left(
\begin{array}
[c]{cccc}%
0 & 0 & 0 & 0\\
0 & b_{1} & b_{2} & 0\\
0 & b_{3} & b_{4} & 0\\
0 & 0 & 0 & 0
\end{array}
\right)  , \label{equ10}%
\end{equation}
where%
\begin{equation}
b_{1}=b_{4}=\frac{1}{1+\left\vert d_{3}\right\vert ^{2}},\text{ }b_{2}%
=\frac{d_{3}^{\ast}}{1+\left\vert d_{3}\right\vert ^{2}},\text{ }b_{3}%
=\frac{d_{3}}{1+\left\vert d_{3}\right\vert ^{2}}, \label{equ11}%
\end{equation}
in which $d_{3}^{\ast}$ is the complex conjugate of $d_{3}$.

In terms of Eq. (\ref{equ6}), the ground-state concurrence is obtained as
$C_{02}=1$. In this case, the concurrence is always the maximum within the
acceptable ranges of parameters and the system is in the Bell state. What's
interesting is that concurrence is independent of the anisotropic parameter,
real and imaginary magnetic fields.

In order to study the properties of entanglement in the $\mathcal{PT}%
$-symmetric broken region \cite{YLee}, Fig. 4 shows that the concurrence of
$\left\vert \varphi_{3}\right\rangle $ varies with $\eta_{0}$. In the
$\mathcal{PT}$-symmetric region, the concurrence is a constant, while it
decreases with the increase of the imaginary magnetic field and tends to zero
infinitely in the $\mathcal{PT}$-symmetric broken region $\left(  \eta
_{0}>1\right)  $. Moreover, the concurrence shows the non-analytic behavior at
the exceptional point.
\begin{figure}
[ptb]
\begin{center}
\includegraphics[
height=4.3007in,
width=5.5884in
]%
{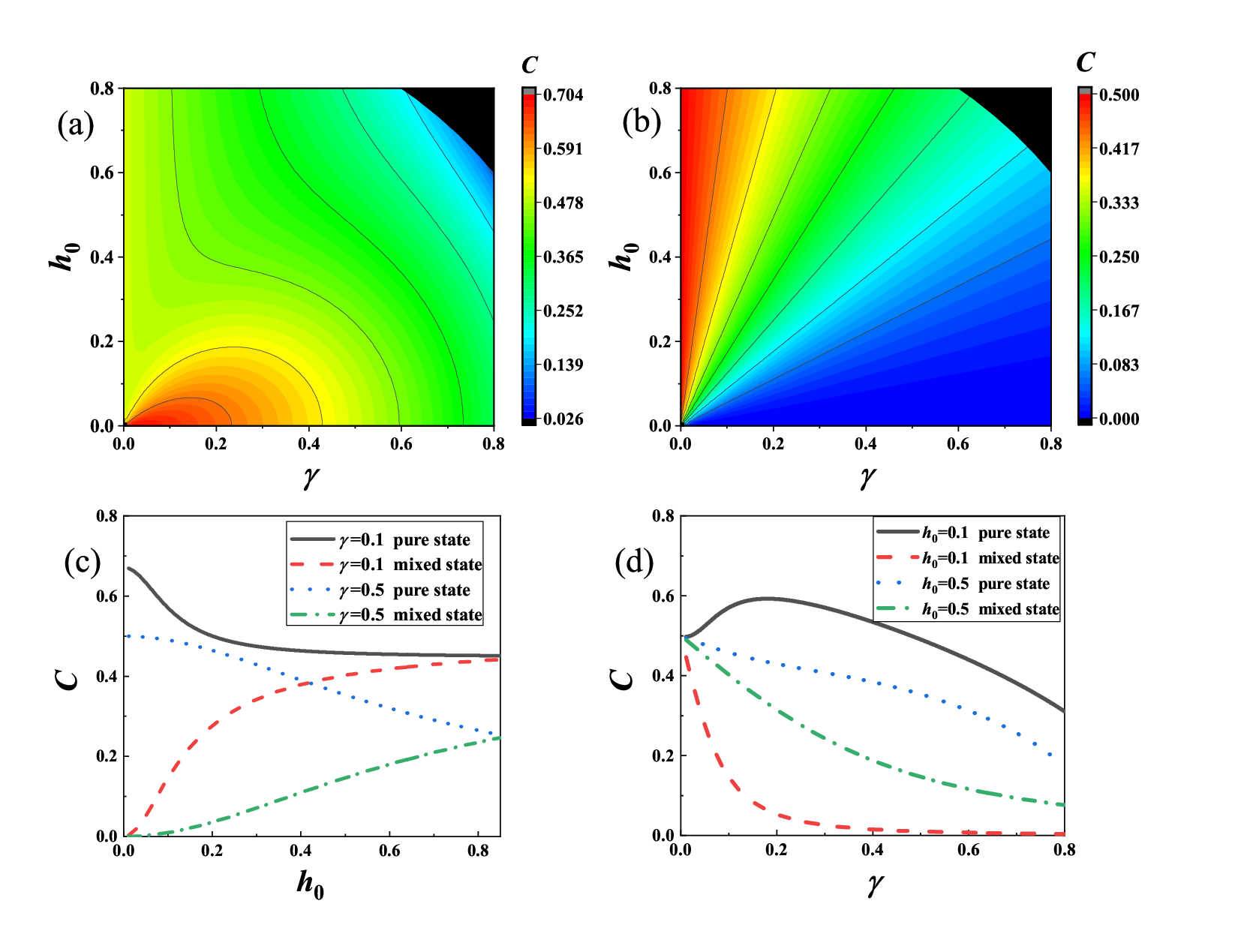}%
\caption{The mixed-state and the pure-state entanglements of two
ground-states. When $\eta_{0}=\sqrt{1-h_{0}^{2}-\gamma^{2}}$, the variations
of concurrence with $h_{0}$ and $\gamma$: $\left(  \text{a}\right)  $ The
contour plot of pure-state entanglement. $\left(  \text{b}\right)  $ The
contour plot of mixed-state entanglement. The black area is a area where
$\gamma$ and $h_{0}$ cannot take a value. $\left(  \text{c}\right)  $ and
$\left(  \text{d}\right)  $ The entanglements of the mixed state and the pure
state as functions of $h_{0}$ when $\gamma=0.1$ and $\gamma=0.5$ and as
functions of $\gamma$ when $h_{0}=0.1$ and $h_{0}=0.5$.}%
\label{Figure5}%
\end{center}
\end{figure}

$\left(  \text{\textbf{3}}\right)  $ \textbf{The case of }$h_{0}^{2}+\eta
_{0}^{2}=1-\gamma^{2}$

On the basis of the Hamiltonian and the above results, it is found that the
entanglement curves are axially symmetric concerning zero. Without loss of
generality, the following discussions only focus on the case where each
parameter is greater than zero. Next, we study the ground-state entanglement
when $\left\vert \varphi_{1}\right\rangle $ and $\left\vert \varphi
_{3}\right\rangle $ are degenerate ground states.

When the superposition of ground states $\left\vert \varphi_{1}\right\rangle $
and $\left\vert \varphi_{3}\right\rangle $ is a pure state, that is,
$\left\vert \varphi_{0}\right\rangle =d_{1}\left\vert \varphi_{1}\right\rangle
+d_{2}\left\vert \varphi_{3}\right\rangle $, where $d_{1}$ and $d_{2}$ are
constants, thus the density matrix is $\rho=\left\vert \varphi_{0}%
\right\rangle \left\langle \varphi_{0}\right\vert $. Moreover, we obtain the
entanglement of the pure state when $d_{1}=d_{2}$.

When the superposition of $\left\vert \varphi_{1}\right\rangle $ and
$\left\vert \varphi_{3}\right\rangle $ is a mixed state, the density matrix is
written as $\rho=\left\vert \varphi_{1}\right\rangle p_{1}\left\langle
\varphi_{1}\right\vert +\left\vert \varphi_{3}\right\rangle p_{2}\left\langle
\varphi_{3}\right\vert $. Assume that $p_{1}=p_{2}$, the mixed-state
entanglement is obtained. \ 

We show the variation of pure-state entanglement with $h_{0}$ and $\gamma$
when $\eta_{0}=\sqrt{1-h_{0}^{2}-\gamma^{2}}(0<h_{0}^{2}+\gamma^{2}\leq1)$ in
Fig. 5$($a$)$ and find that the concurrence decreases with the increasing
$h_{0}$, which is opposite of the entanglement of the mixed state in Fig.
5$($b$)$. Fig. 5$($c$)$ is the variations of the pure-state entanglement and
the mixed-state one with $h_{0}$ when $\gamma=0$ and $\gamma=0.5$, and we find
that the former is greater than the latter. Furthermore, the pure-state
entanglement first increases and then decreases with the increasing $\gamma$
when $h_{0}=0.1$ in Fig. 5$($d$)$, which is also consistent with the change
rule of Fig. 5$($a$)$.%
\begin{figure}
[ptb]
\begin{center}
\includegraphics[
height=3.7775in,
width=5.047in
]%
{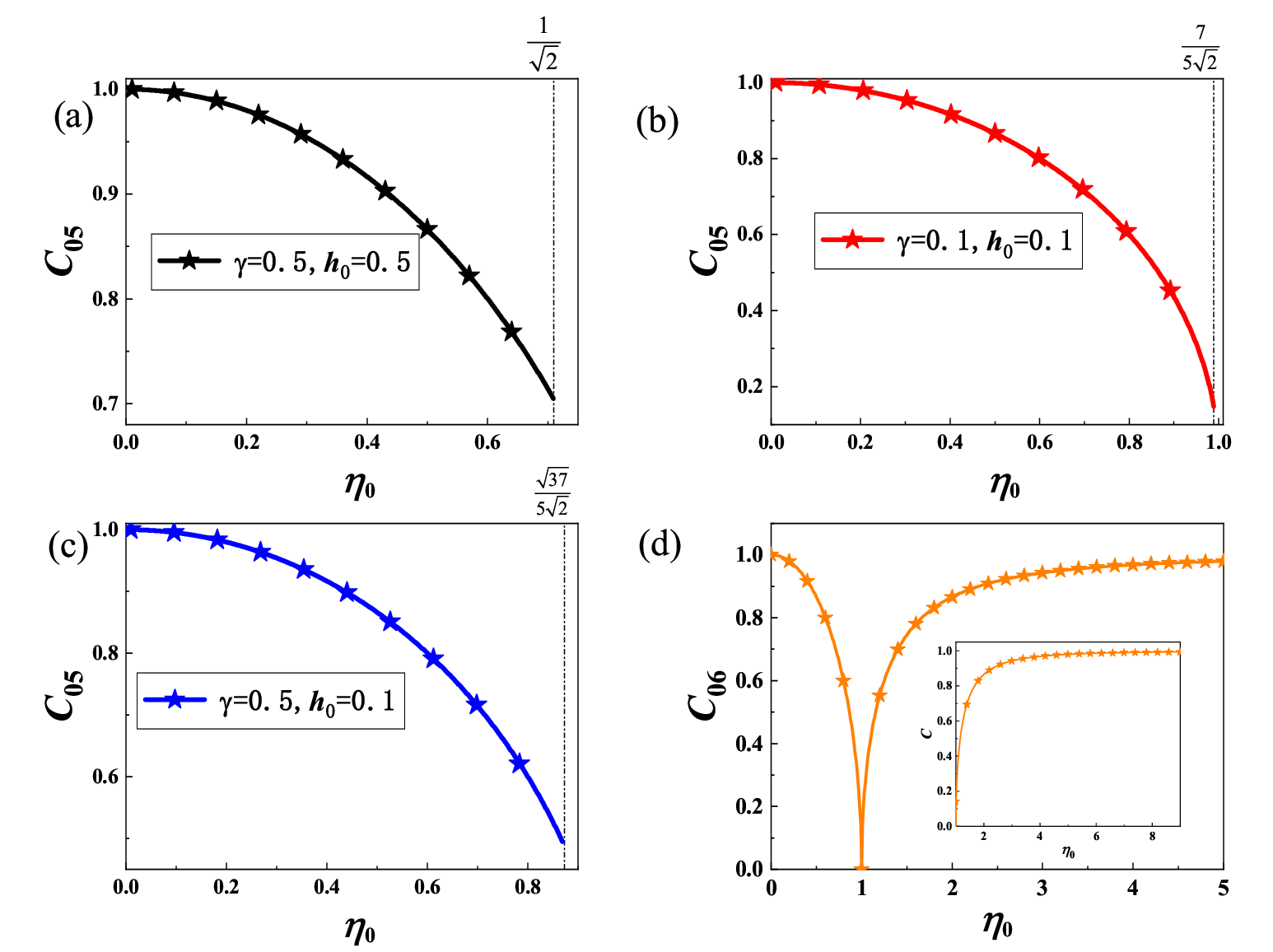}%
\caption{Concurrence versus imaginary magnetic field. (a), (b) and (c) are the
change of ground-state concurrence with $\eta_{0}$ for different $\gamma$ and
$h_{0}$ in the $\mathcal{PT}$-symmetry region. The range of $\eta_{0}$ between
zero and the marked dashed line can take values. (d) The concurrence of
$\left\vert \varphi_{3}\right\rangle $. The illustration shows concurrence
when the system is in the $\mathcal{PT}$-symmetry broken region.}%
\label{Fig.6}%
\end{center}
\end{figure}

\section{Bi-orthogonal Basis\label{sec50}}

Due to non-Hermitian Hamiltonian $H\neq H^{\dagger},$ the eigenvalue equations
of $H$ and $H^{\dagger}$ are given by \cite{57,58}%
\begin{equation}
H\left\vert \varphi_{n}\right\rangle =E_{n}\left\vert \varphi_{n}\right\rangle
,\text{ }\left\langle \varphi_{n}\right\vert H^{\dagger}=\left\langle
\varphi_{n}\right\vert E_{n}^{\ast} \label{equ12}%
\end{equation}%
\begin{equation}
H^{\dagger}\left\vert \phi_{m}\right\rangle =E_{m}^{^{\prime}}\left\vert
\phi_{m}\right\rangle ,\text{ }\left\langle \phi_{m}\right\vert H=\left\langle
\phi_{m}\right\vert E_{m}^{^{\prime}\ast} \label{equ13}%
\end{equation}
where%
\begin{equation}
\left\langle \phi_{m}\right\vert \varphi_{n}\rangle=\delta_{mn}. \label{equ14}%
\end{equation}
On the basis, the biorthogonal density matrix can be written as%
\begin{equation}
\rho=\left\vert \varphi_{n}\right\rangle \left\langle \phi_{n}\right\vert .
\label{equ15}%
\end{equation}
The eigenvalues and corresponding eigenstates of $H^{\dagger}$ are obtained
as
\begin{subequations}
\begin{align}
E_{1}^{^{\prime}}  &  =-2J\sqrt{h_{0}^{2}+\gamma^{2}},\text{ }\left\vert
\phi_{1}\right\rangle =\frac{1}{\sqrt{\gamma^{2}+d_{1}^{2}}}\left[
d_{1}\left\vert \uparrow\uparrow\right\rangle +\gamma\left\vert
\downdownarrows\right\rangle \right]  ,\label{equ16a}\\
E_{2}^{^{\prime}}  &  =2J\sqrt{h_{0}^{2}+\gamma^{2}},\text{ }\left\vert
\phi_{2}\right\rangle =\frac{1}{\sqrt{\gamma^{2}+d_{2}^{2}}}\left[
d_{2}\left\vert \uparrow\uparrow\right\rangle +\gamma\left\vert
\downdownarrows\right\rangle \right]  ,\label{equ16b}\\
E_{3}^{^{\prime}}  &  =-2J\sqrt{1-\eta_{0}^{2}},\text{ }\left\vert \phi
_{3}\right\rangle =\frac{1}{\sqrt{1+\left\vert d_{3}^{\ast}\right\vert ^{2}}%
}\left[  d_{3}^{\ast}\left\vert \uparrow\downarrow\right\rangle +\left\vert
\downarrow\uparrow\right\rangle \right]  ,\label{equ16c}\\
E_{4}^{^{\prime}}  &  =2J\sqrt{1-\eta_{0}^{2}},\text{ }\left\vert \phi
_{4}\right\rangle =\frac{1}{\sqrt{1+\left\vert d_{4}^{\ast}\right\vert ^{2}}%
}\left[  d_{4}^{\ast}\left\vert \uparrow\downarrow\right\rangle +\left\vert
\downarrow\uparrow\right\rangle \right]  . \label{equ16d}%
\end{align}

When $\left\vert \varphi_{3}\right\rangle $ is the ground state, the
biorthogonal density matrix is%
\end{subequations}
\begin{equation}
\rho_{03}=\left(
\begin{array}
[c]{cccc}%
0 & 0 & 0 & 0\\
0 & x_{1} & x_{2} & 0\\
0 & x_{3} & x_{4} & 0\\
0 & 0 & 0 & 0
\end{array}
\right)  , \label{equ17}%
\end{equation}
where%
\begin{align}
x_{1}  &  =\frac{d_{3}^{2}}{\sqrt{\left(  1+\left\vert d_{3}\right\vert
\right)  \left(  1+\left\vert d_{3}^{\ast}\right\vert \right)  }},\text{
}x_{2}=\frac{d_{3}}{\sqrt{\left(  1+\left\vert d_{3}\right\vert \right)
\left(  1+\left\vert d_{3}^{\ast}\right\vert \right)  }},\nonumber\\
x_{3}  &  =\frac{d_{3}}{\sqrt{\left(  1+\left\vert d_{3}\right\vert \right)
\left(  1+\left\vert d_{3}^{\ast}\right\vert \right)  }},\text{ }x_{4}%
=\frac{1}{\sqrt{\left(  1+\left\vert d_{3}\right\vert \right)  \left(
1+\left\vert d_{3}^{\ast}\right\vert \right)  }}. \label{equ18}%
\end{align}

According to the definition Eq. (\ref{equ6}), the ground-state concurrence of
the system can be obtained in the case of biorthogonal basis in Figs.
6(a)$-$(c), and it decreases with increasing $\eta_{0}$. In this case, the
entanglement of this non-Hermitian system is smaller than that of the
Hermitian system $\left(  \eta_{0}=0\right)  $, which is different from the
case of the density matrix of $\left\vert \varphi_{3}\right\rangle $ in the
region of PT symmetry. When $\left\vert \varphi_{1}\right\rangle $ is the
ground state, the concurrence is identical with the case of $\rho_{01}$ due to
$\left\vert \varphi_{1}\right\rangle =\left\vert \phi_{1}\right\rangle $.

The variations of concurrence in biorthogonal basis of $\left\vert \varphi
_{3}\right\rangle $ are shown in Fig. 6(d). It is found that the trends of
entanglement in the $\mathcal{PT}$-symmetry region and broken region are
inverse. At the exceptional point, the entanglement is reduced to the minimum,
which is the opposite of the result in Fig. 4, and has the non-analytic
behavior.
\begin{figure}
[ptb]
\begin{center}
\includegraphics[
height=3.6737in,
width=5.047in
]%
{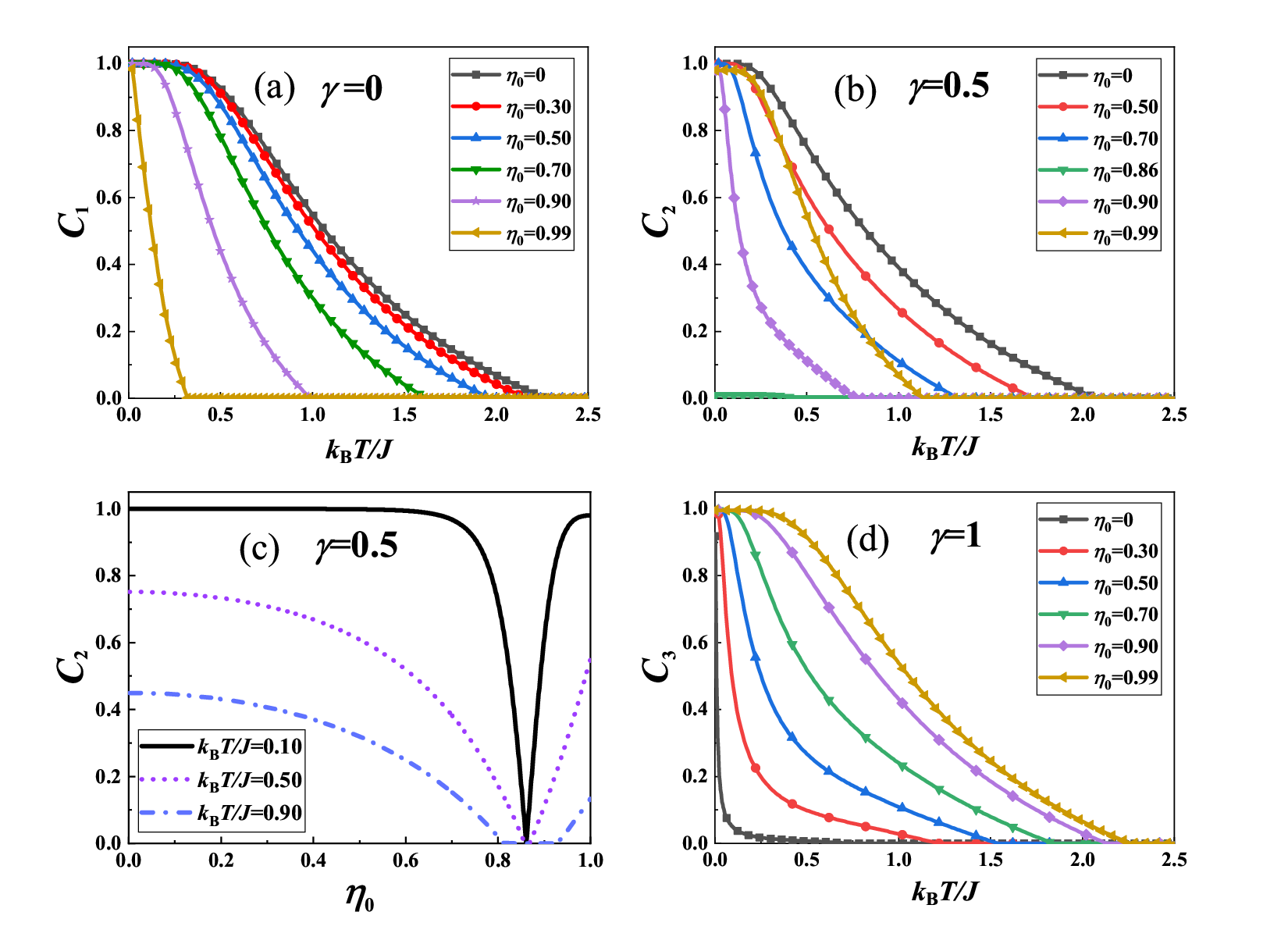}%
\caption{The thermal entanglements between two spins for different values of
the anisotropy parameters and imaginary magnetic fields when $h_{0}=0.1$: (a)
$\gamma=0$, entanglement is the maximum when $\eta_{0}=0$. (b) $\gamma=0.5$,
the thermal entanglement as a function of temperature. (c) $\gamma=0.5$, the
thermal entanglement as functions of $\eta_{0}$ for different values of
temperature. (d) $\gamma=1$, entanglement is the minimum when $\eta_{0}=0$. }%
\label{Figure7}%
\end{center}
\end{figure}
\begin{figure}
[ptb]
\begin{center}
\includegraphics[
height=5.2191in,
width=5.047in
]%
{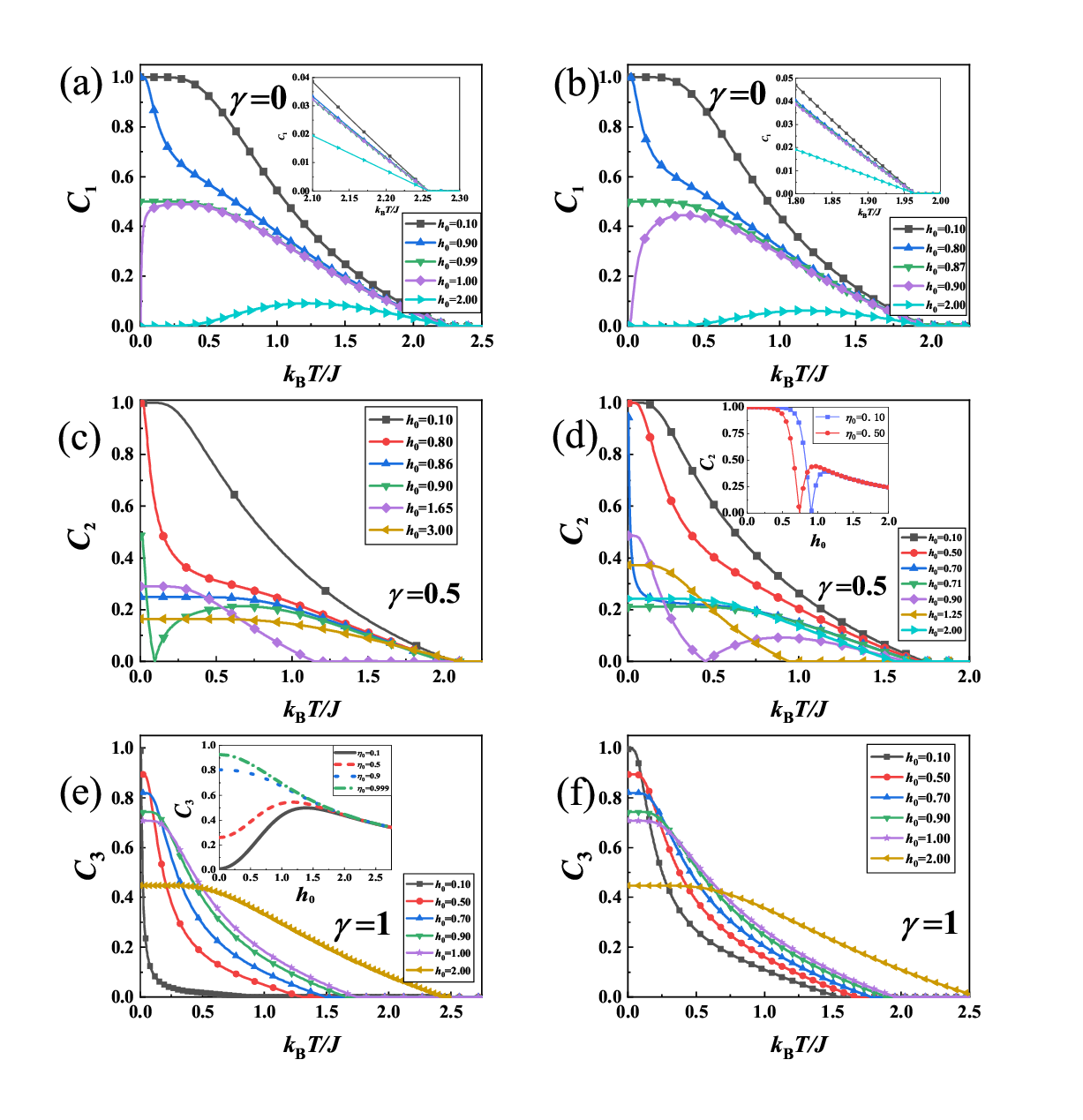}%
\caption{The thermal entanglements between two spins as the functions of
temperature for different values of real and imaginary magnetic fields when
$\gamma=0$, $0.5$ and 1. (a), (c) and (e) are $\eta_{0}=0.1$ and (b), (d) and
(f) are $\eta_{0}=0.5$. (a) the illustration shows the behavior of concurrence
at a temperature around 2.1 to 2.3. (b) the illustration shows the behavior of
concurrence at a temperature around 1.8 to 2. (d) the illustration shows
variation of concurrence with $h_{0}$ when $\gamma=0.5$, $k_{\text{B}}%
T/J=0.1$. (e) the illustration shows variation of concurrence with $h_{0}$
when $\gamma=1$, $k_{\text{B}}T/J=0.5$.}%
\end{center}
\end{figure}

\section{Thermal Entanglement\label{sec5}}

In this section, the thermal entanglement of the system is studied in the
$\mathcal{PT}$-symmetry region. The density matrix is defined as $\rho
=\exp\left(  -\beta H\right)  \diagup Z$, in which $\beta=1\diagup
k_{\text{B}}T$, $k_{\text{B}}$ is the Boltzmann constant, $T$ is temperature,
and $Z=$Tr$\left[  \exp\left(  -\beta H\right)  \right]  $ is the partition
function in the canonical ensemble. In this way, we obtain the density matrix
and concurrence of the system.%
\begin{figure}
[ptb]
\begin{center}
\includegraphics[
height=3.6737in,
width=5.047in
]%
{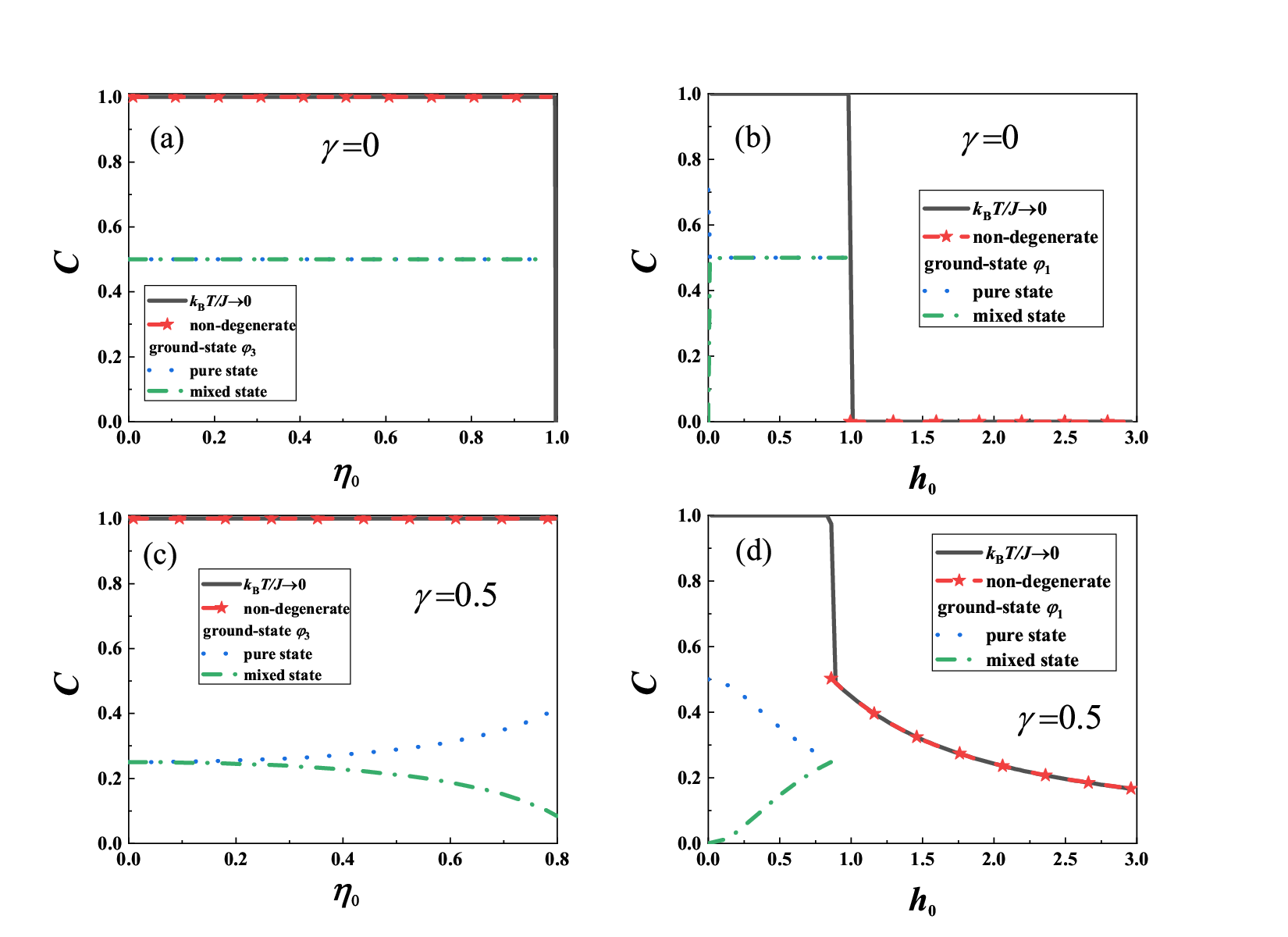}%
\caption{The comparison between thermal entanglement as $k_{\text{B}%
}T/J\rightarrow0,$ the pure-state, mixed-state and non-degenerate ground-state
entanglements. (a) and (c): Black (Solid) and red (star) lines correspond to
$h_{0}=0.1$. Blue (dotted) and green (chain) lines correspond to $h_{0}%
=\sqrt{1-\eta_{0}^{2}-\gamma^{2}}$. (b) and (d): Black (Solid) and red (star)
lines correspond to $\eta_{0}=0.1$. Dotted (blue) and chain (green) lines
correspond to $\eta_{0}=\sqrt{1-h_{0}^{2}-\gamma^{2}}$.}%
\end{center}
\end{figure}

Fig. 7 shows the variations of the concurrence with temperature for different
$\eta_{0}$ and $\gamma$. We find that the concurrence decreases with the
increase of $\eta_{0}$ at the same temperature when $\gamma=0$, and the larger
$\eta_{0}$ is, the faster the concurrence decreases. When $\gamma=0.5$, the
concurrence first decreases then increases with the increase of $\eta_{0}$,
which can be seen more intuitively in Fig. 7(c), and it reduces to a minimum
when $\eta_{0}=0.86$ due to the ground state is degenerate in this point. For
$\gamma=1$, the concurrence increases with the increase of $\eta_{0}$, and it
indicates that the entanglement of this non-Hermitian system is greater than
that of the Hermitian system $\left(  \eta_{0}=0\right)  $, which is opposite
of the case when $\gamma=0$.

Fig. 8$\left(  \text{a}\right)  $ shows the variations of the concurrence with
temperature for different $h_{0}$ and $\eta_{0}$ when $\gamma=0$. Due to the
energy level crossing, the concurrence can change suddenly when the magnetic
field near $0.99$ as $k_{\text{B}}T/J\rightarrow0$. In addition, it decreases
with the increase of $h_{0}$ at the same temperature when $\eta_{0}=0.1$ and
reduces directly to zero with the increase of temperature when $0\leq
h_{0}\leq0.99$. It is worth noticing that the concurrence first increases and
then decreases with the increasing temperature when $h_{0}>0.99$. We can
obviously see that the concurrence under different external magnetic fields
becomes zero at the same temperature in the illustration, which indicates that
the threshold temperature is independent of $h_{0}$. When $\eta_{0}=0.5$, the
variations of concurrence with temperature are shown in Fig. 8$\left(
\text{b}\right)  $, which is similar to the case when $\eta_{0}=0.1$, and the
point of sudden change is in $h_{0}=0.87$. The concurrence decreases faster
with the increase of temperature, and the temperature at which eventually
becomes zero is smaller than that in Fig. 8$($a$)$, which is consistent with
the result when $\gamma=0$ in Fig. 7. Furthermore, we find that the larger
$\eta_{0}$ is, the smaller $h_{0}$ corresponding to the point of sudden change
of concurrence.

In Fig. 8(c), the variations of the concurrence with temperature are shown for
different $h_{0}$ and $\eta_{0}$ when $\gamma=0.5$. For $\eta_{0}=0.1$, we
find that the concurrence reduces directly to zero with the increasing
temperature when $0\leq h_{0}\leq0.86$. When $h_{0}>0.86$, the concurrence
first decreases, then increases, and then decreases to zero with the increase
of temperature. When $h_{0}$ increases to around $1.65$, the concurrence
decreases directly to zero with the increase of temperature. The concurrence
first decreases, then increases, and then decreases with the increase of
$h_{0}$ when $k_{\text{B}}T/J\rightarrow0$, and this is more evident in the
inset of Fig. 8$($d$)$. For $\eta_{0}=0.5$, the concurrence is similar to the
case when $\eta_{0}=0.1$. The point of sudden change is obtained in
$h_{0}=0.71$. The temperature at which the concurrence eventually becomes zero
is smaller, and it is consistent with the result when $\gamma=0.5$ in Fig. 7.

Fig. 8$\left(  \text{e}\right)  $ and Fig. 8$\left(  \text{f}\right)  $ show
the variations of concurrence with temperature for different $h_{0}$ and
$\eta_{0}$ when $\gamma=1$, respectively. It is observed that the concurrence
decreases with the increase of $h_{0}$ when $k_{\text{B}}T/J\rightarrow0$,
while the real magnetic field enhances entanglement when the temperature
becomes higher. It is also found that the smaller $h_{0}$ is, the faster the
concurrence decreases. When $h_{0}=2$, the variation of concurrence with
temperature for $\eta_{0}=0.1$ is identical to that of $\eta_{0}=0.5$, as a
consequence the concurrence is independent of $\eta_{0}$ in this case, which
can be seen more distinctly from the illustration in Fig. 8$\left(
\text{e}\right)  $. The concurrence decreases smoothly to zero with the
increasing temperature when $\gamma=1$, which is different from $\gamma=0$ and
$\gamma=0.5$, due to the system becomes Ising model at this moment.

In order to compare the thermal entanglement as $k_{\text{B}}T/J\rightarrow0$,
the pure-state, mixed-state and non-degenerate ground-state entanglements,
Fig. 9 shows their variations with $h_{0}$ and $\eta_{0}$. It can be seen
that, within the ranges of available values, the pure-state and mixed-state
entanglements are overlapped partially. When $k_{\text{B}}T/J\rightarrow0,$
the thermal and the non-degenerate ground-state entanglements have overlapping
parts. It is indicated that the thermal entanglement is realized by
non-degenerate ground states in this case.%
\begin{figure}
[ptb]
\begin{center}
\includegraphics[
height=3.9029in,
width=5.047in
]%
{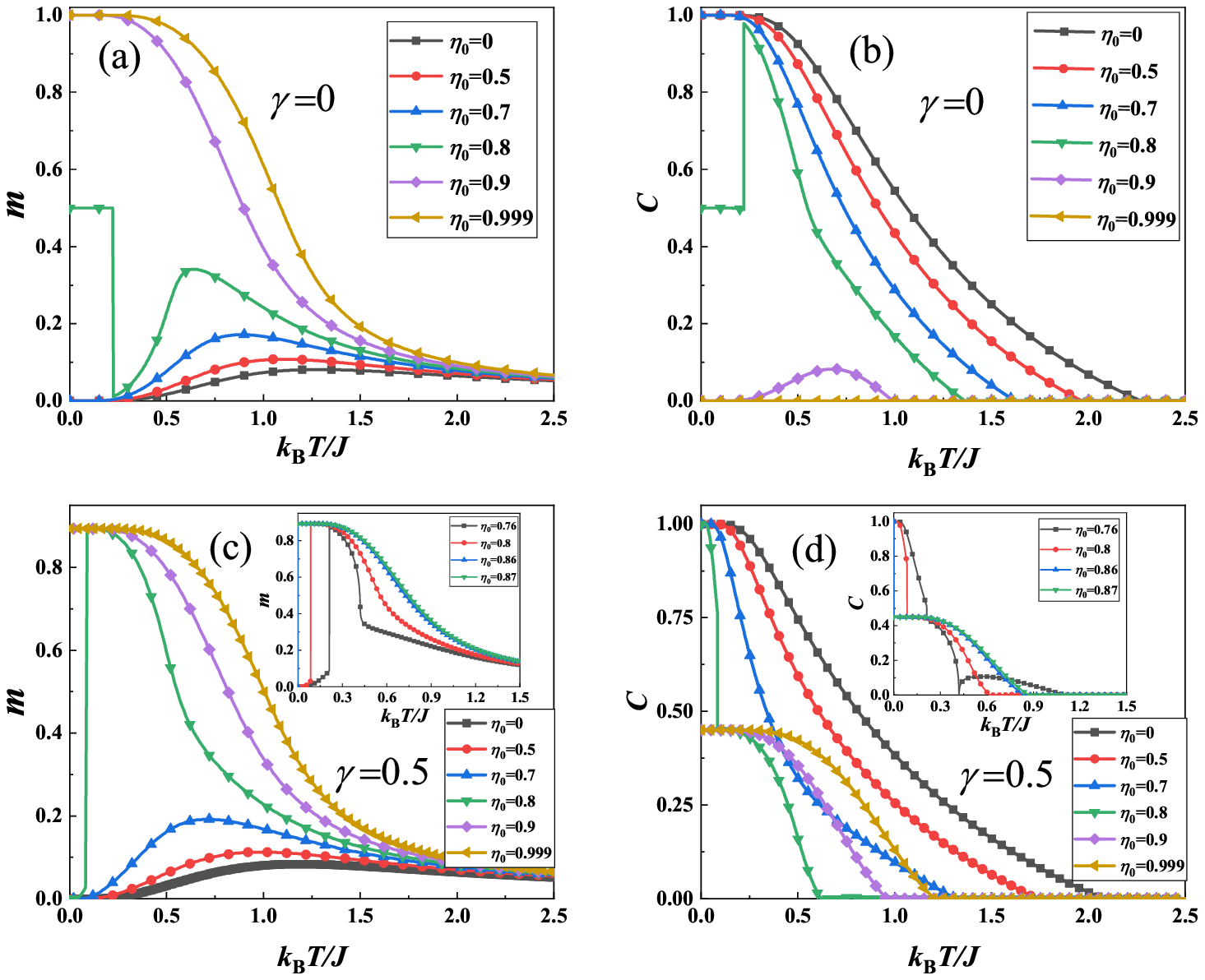}%
\caption{The magnetization and concurrence as the functions of temperature for
several different the imaginary magnetic fields when $\gamma=0,0.5$ and
$h_{0}=0.1$. They have the opposite change\ trends, that is, the magnetization
increases when the concurrence decreases, and change suddenly when $\eta
_{0}=0.8$. The insets of (c) and (d) show the variations of the magnetization
and the concurrence with temperature around $\eta_{0}=0.8$. }%
\end{center}
\end{figure}
\begin{figure}
[ptb]
\begin{center}
\includegraphics[
height=2.2035in,
width=5.5486in
]%
{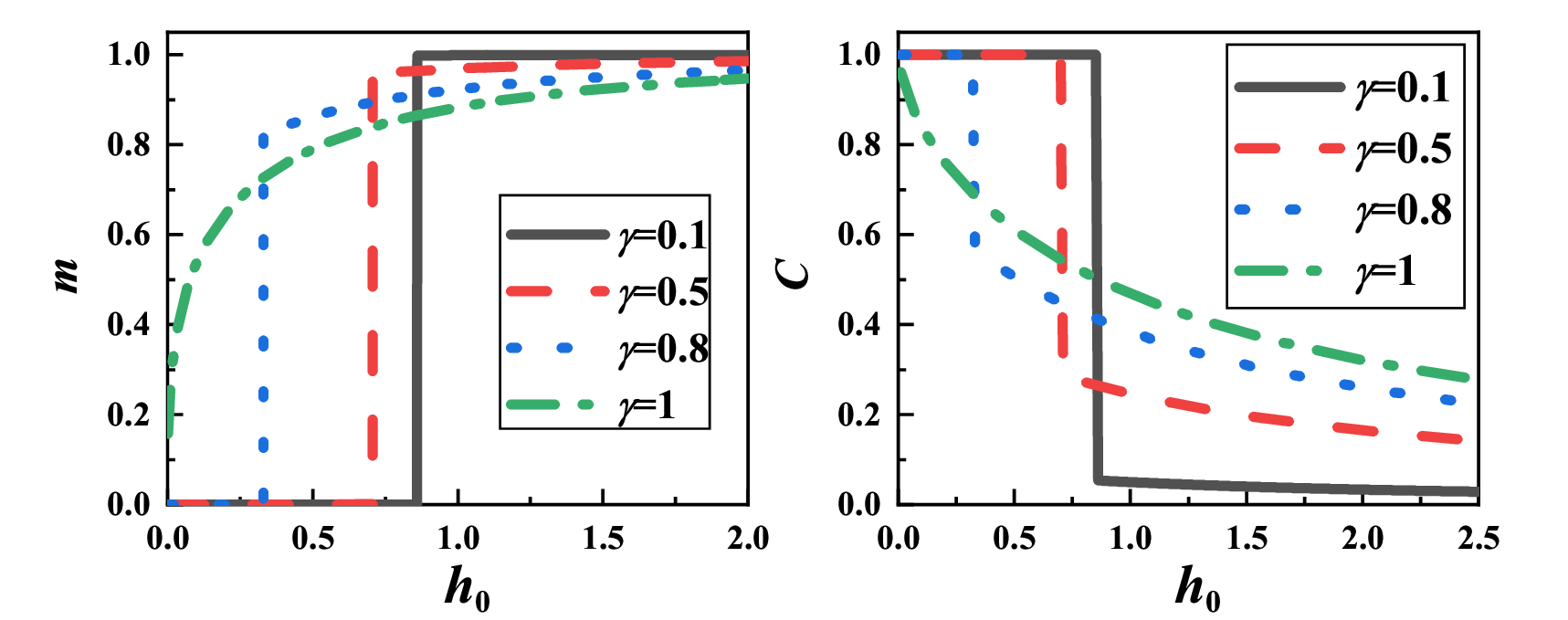}%
\caption{The magnetization and concurrence as the functions of the magnetic
field for several different the anisotropy parameters as $k_{\text{B}%
}T/J\rightarrow0$. They have the opposite change trends, and their points of
sudden change are the same. }%
\end{center}
\end{figure}

\section{Result of Mean-Field Theory\label{sec6}}

Based on the previous discussions of two-site spin system, we study the
magnetization and entanglement of the non-Hermitian spin-1/2 XY spin chain by
using the mean-field theory in this section. According to the two-spin cluster
mean-field approximation \cite{59,60}, the many-body system is transformed
into the two-body system, and the two-spin cluster Hamiltonian can be written
as%
\begin{align}
H_{MFA}  &  =-\frac{J}{2}\left[  \left(  1+\gamma\right)  (\sigma_{1}%
^{x}\sigma_{2}^{x}+\sigma_{2}^{x}\sigma_{1}^{x})+\left(  1-\gamma\right)
(\sigma_{1}^{y}\sigma_{2}^{y}+\sigma_{2}^{y}\sigma_{1}^{y})\right]
\label{equ20}\\
&  -h(\sigma_{1}^{z}+\sigma_{2}^{z})+i\eta(-\sigma_{1}^{z}+\sigma_{2}%
^{z})-Jm(q-1)(\sigma_{1}^{z}+\sigma_{2}^{z}),\nonumber
\end{align}
where $m=\left\langle \frac{1}{2}\left(  \sigma_{1}^{z}+\sigma_{2}^{z}\right)
\right\rangle $ is the magnetization along the fixed direction $z$ in space.
In terms of above method, the average effective Hamiltonian of two-spin
cluster is%
\begin{align}
\widetilde{H}_{MFA}  &  =-\frac{J}{2}\left[  \left(  1+\gamma\right)
(\sigma_{1}^{x}\sigma_{2}^{x}+\sigma_{2}^{x}\sigma_{1}^{x})+\left(
1-\gamma\right)  (\sigma_{1}^{y}\sigma_{2}^{y}+\sigma_{2}^{y}\sigma_{1}%
^{y})\right]  -h(\sigma_{1}^{z}+\sigma_{2}^{z})\label{equ21}\\
&  +i\eta(-\sigma_{1}^{z}+\sigma_{2}^{z})-Jm(q-1)(\sigma_{1}^{z}+\sigma
_{2}^{z})+J(q-1)m^{2},\nonumber
\end{align}
and the free energy as a function of the temperature $T$ and magnetization $m$
is given by%
\begin{equation}
F=-k_{B}T\ln\widetilde{Z}=J\left(  q-1\right)  m^{2}-k_{B}T\ln Z,
\label{equ22}%
\end{equation}
where $\widetilde{Z}=$Tr$\left[  \exp\left(  -\beta\widetilde{H}_{MFA}\right)
\right]  $ and $Z=$Tr$\left[  \exp\left(  -\beta H_{MFA}\right)  \right]  $.
By the equilibrium condition $\partial F/\partial m=0$ of the system, the
magnetization $m$ is obtained as%
\begin{equation}
m=\frac{\left[  h_{0}+m\left(  q-1\right)  \right]  \sinh\left(  \frac
{2J\sqrt{\left[  h_{0}+m\left(  q-1\right)  \right]  ^{2}+\gamma^{2}}}{k_{B}%
T}\right)  }{\sqrt{\left[  h_{0}+m\left(  q-1\right)  \right]  ^{2}+\gamma
^{2}}\left[  \cosh\left(  \frac{2J\sqrt{\left[  h_{0}+m\left(  q-1\right)
\right]  ^{2}+\gamma^{2}}}{k_{B}T}\right)  +\cosh\left(  \frac{2J\sqrt
{1-\eta^{2}}}{k_{B}T}\right)  \right]  }. \label{equ23}%
\end{equation}%
\begin{figure}
[ptb]
\begin{center}
\includegraphics[
height=4.7115in,
width=5.047in
]%
{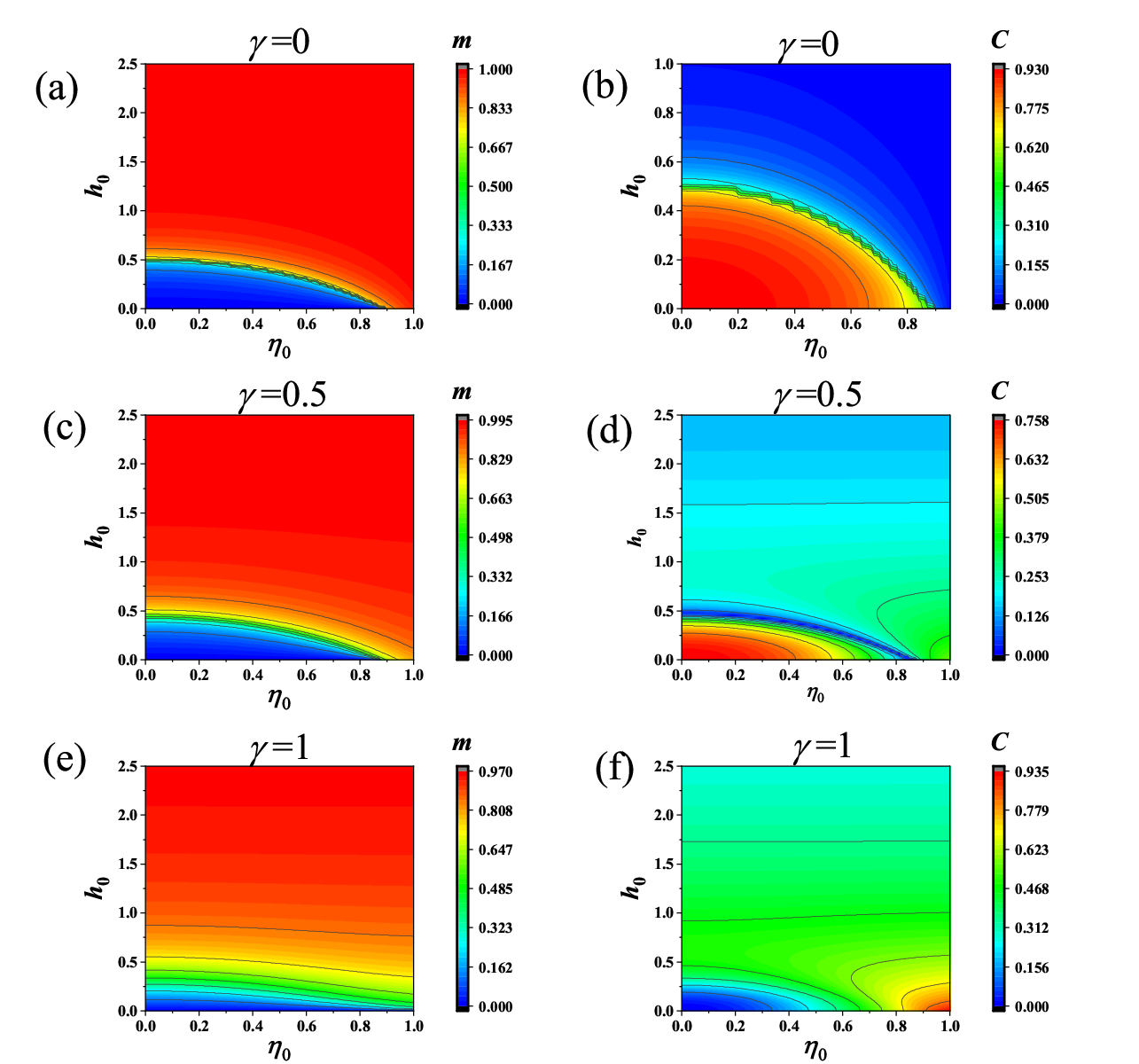}%
\caption{The contour plots of the magnetization and the concurrence for three
different the anisotropy parameters when $k_{\text{B}}T/J=0.5$. (a), (c) and
(e) are the contour plots of the magnetization. (b), (d) and (f) are the
contour plots of the concurrence. }%
\end{center}
\end{figure}

On the basis of the previous discussion, without loss of generality, we only
calculate a few outstanding results. In Fig. 10, we obtain the variations of
the magnetization and the concurrence with temperature for different $\eta
_{0}$ when $\gamma=0,0.5$ and $h_{0}=0.1$ and find that the magnetization
increases and the concurrence decreases with the increase of $\eta_{0}$ at the
same temperature in Figs. 10(a) and (b), furthermore, their variations have
opposite trends with temperature, which is caused by two-spin cluster
mean-field approximation used in Eq. (\ref{equ21}). When $\eta_{0}=0.8$, the
magnetization and entanglement become 0.5 at $k_{\text{B}}T/J=0$ due to the
ground states are degenerate, and their sudden change which is the result of
the self-consistent equation Eq. (\ref{equ23}), which is different from the
result of\ the two-site system. In addition, the concurrence is maximum when
$\eta_{0}=0$, which implies that the entanglement of this non-Hermitian system
is smaller than one of the Hermitian system $\left(  \eta_{0}=0\right)  $.
This is similar to the case of the system of two sites. From Figs. 10(c) and
(d), the magnetization and the concurrence change suddenly as $\eta_{0}$
increases to around $0.8$, which can be seen more clearly in the
illustrations, and the maximum value of entanglement becomes smaller when
$\eta_{0}>0.86$ and $k_{\text{B}}T/J=0$. Furthermore, the temperature at which
entanglement finally reduces to zero decreases and then increases with the
increase of $\eta_{0}$.%
\begin{figure}
[ptb]
\begin{center}
\includegraphics[
height=3.5336in,
width=5.047in
]%
{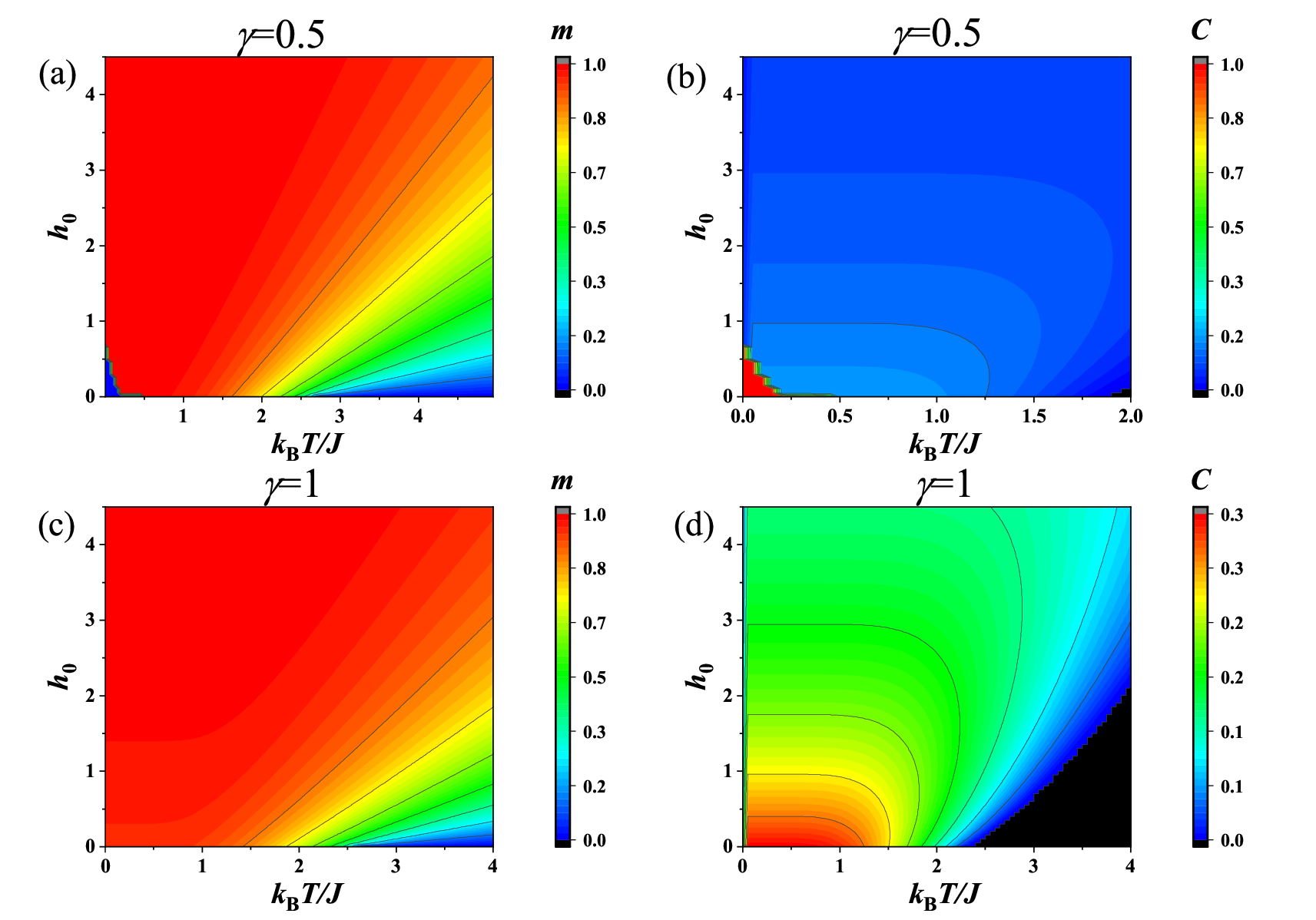}%
\caption{The contour plots of the magnetization and the concurrence with the
real magnetic field and temperature when $q=4$. (a) and (c) are the contour
plots of the magnetization. (b) and (d) are the contour plots of the
concurrence.}%
\end{center}
\end{figure}

In Fig. 11, the magnetization and the concurrence are presented as functions
of $h_{0}$ for four different $\gamma$ when $\eta_{0}=0.5$ and $k_{\text{B}%
}T/J\rightarrow0$, and their variations have opposite trends with the magnetic
field. As we can see in the figure, the discontinuities in the magnetization
and the entanglement occur when $\gamma<1$ in which the system undergoes
first-order quantum phase transitions. Fig. 12 is the contour plots of the
magnetization and the concurrence with $h_{0}$ and $\eta_{0}$ for three
different $\gamma$ when $k_{\text{B}}T/J=0.5$. The magnetization increases
with the increase of the magnetic field, and the concurrence is consistent
with the previous analysis in Fig. 10 for the same case. Especially, the
entanglement first decreases and then increases with $\eta_{0}$ when
$\gamma=0.5$ and $h_{0}$ is small. In addition, the changes of the
magnetization and the concurrence have the same trends when $\gamma=1$, which
is different from the case of $k_{\text{B}}T/J\rightarrow0$ in Fig. 11.

The properties of the one-dimensional systems have been introduced in the
previous sections, and the high-dimensional systems also can be discussed by
this mean-field method. In Fig. 13, the contour plots of the magnetization and
the concurrence with $h_{0}$ and $k_{\text{B}}T/J$ for two different $\gamma$
when $\eta_{0}=0.5$ in the two-dimensional $\left(  q=4\right)  $ system are
shown. We find that it is similar to the one-dimensional case but the
temperature at which the concurrence finally reaches zero is higher in the
two-dimensional case than in the one-dimensional case and the entanglement of
the three-dimensional system also is similar, whereas aforementioned the
temperature is higher.

\section{Conclusions\label{sec7}}

In this manuscript, we have studied the ground-state and thermal entanglements
of the non-Hermitian spin-1/2 XY model. In$\ \mathcal{PT}$-symmetric region,
it is found that the two-site entanglement of $\left\vert \varphi
_{3}\right\rangle $\ is independent of the imaginary magnetic field $\eta_{0}%
$, while $\eta_{0}$ weakens\ the entanglement$\ $for the case of the
biorthogonal basis in the two-site system. Moreover, the concurrence of
$\left\vert \varphi_{3}\right\rangle $ shows the non-analytic behavior at the
exceptional point, and it indicates that the concurrence can characterize the
phase transition in this non-Hermitian system. In addition, there are the
first-order quantum phase transitions in this one-dimensional chain for some
anisotropic parameters\ in the region of $\mathcal{PT}$ symmetry, and the
entanglement changes suddenly at the quantum phase transition point. For
thermal\ entanglement, the imaginary magnetic field weakens it when the system
is isotropic and enhances it when the anisotropy parameter $\gamma=1$ $\left(
\text{Ising model}\right)  $.

\begin{acknowledgments}
This work is supported by the National Natural Science Foundation of China
under Grants No. 11675090, and No. 11905095; Shandong Provincial Natural
Science Foundation, China under Grant No. ZR202111160185. Y. Li. would like to
thank Chun-Yang Wang, Jing Wang, Zhen-Hui Sun, Xiu-Ying Zhang, Qing-Hui Li and
Chuan-Zheng Miao for fruitful discussions and useful comments.
\end{acknowledgments}

\textbf{Data availability }All data generated or analyzed during this study
are included in this published article.

\bigskip\textbf{Declarations}

\textbf{Conflict of interest} The authors have no known competing financial
interests or personal relationships that could have appeared to influence the
work reported in this paper.

\end{document}